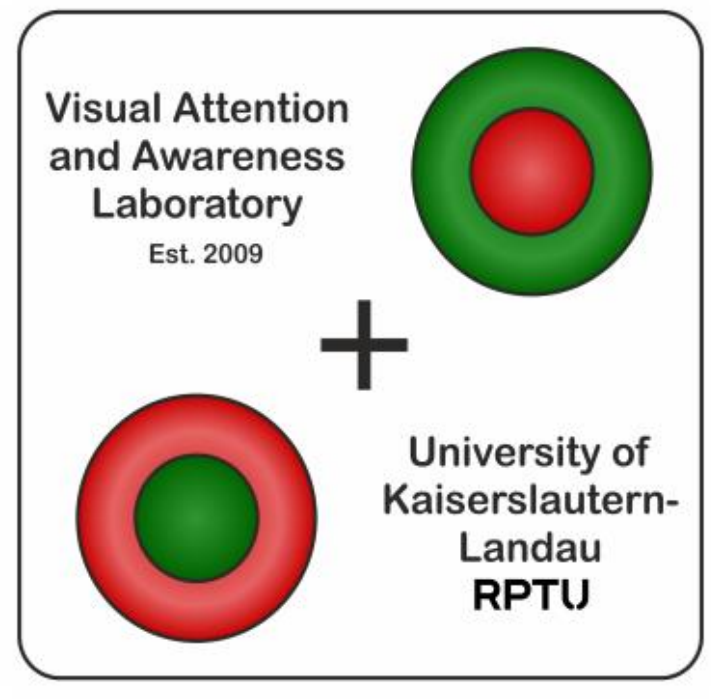

# Functional dissociations versus post-hoc selection: Moving beyond the Stockart et al. (2025) compromise

**Thomas Schmidt, Xin Ying Lee, & Maximilian P. Wolkersdorfer**

Center for Cognitive Science

University of Kaiserslautern-Landau (RPTU), Germany

Corresponding author:

Thomas Schmidt
University of Kaiserslautern-Landau (RPTU), Center for Cognitive Science, Visual Attention & Awareness Laboratory
Erwin-Schrödinger-Str. Geb. 57
D-67663 Kaiserslautern, Germany
E-Mail: thomas.schmidt@rptu.de

## Abstract

**Stockart et al. (2025) recommend guidelines for best practices in the field of unconscious cognition. However, they condone the repeatedly criticized technique of excluding trials with high visibility ratings or of participants with high sensitivity for the critical stimulus. Based on standard signal detection theory for discrimination judgments, we show that post-hoc trial selection only isolates points of neutral response bias but remains consistent with uncomfortably high levels of sensitivity. We argue that post-hoc selection constitutes a sampling fallacy that capitalizes on chance, generates regression artifacts, and wrongly ascribes unconscious processing to stimulus conditions that are far from indiscriminable. As an alternative, we advocate the study of functional dissociations, where direct (*D*) and indirect (*I*) measures are conceptualized as spanning up a two-dimensional *D-I* space and where single, sensitivity, and double dissociations appear as distinct curve patterns. While Stockart et al.'s recommendations cover only a single line of that space where $D \approx 0$, functional dissociations can utilize the entire space, circumventing requirements like null visibility, and exhaustive reliability and allowing for the planful measurement of theoretically meaningful functional relationships between experimentally controlled variables.**

## 1. Dissociation curves: A journey through D-I space

Stockart et al. (2025) present the outcome of a number of discussion meetings on "best practices" in the methodology of unconscious cognition research. The result may be viewed as a compromise between a majority recommending a practice called *post-hoc selection* and a minority group strongly criticizing this practice on mathematical grounds. Here we will argue that the compromise is not viable and that post-hoc selection needs to be abandoned because it fails to isolate stimulus conditions where perceptual sensitivity is low. At the same time, we want to point out that there are many attractive alternatives to post-hoc selection that hold great promise for advancing the field. Therefore, this paper should be viewed not only as a critique of post-hoc selection, but also as an invitation to the fascinating field of *functional dissociations* in the study of unconscious cognition.

What are functional dissociations, and what is special about them? Stockart et al.'s paper is anchored in the *dissociation paradigm*, which consists in comparing a direct measure (*D*) of awareness with an indirect measure (*I*) that implies that the critical stimulus was cognitively processed (Reingold & Merikle, 1988; see also Cheesman & Merikle, 1984, 1986; Erdelyi, 1986; Snodgrass et al., 2004). For instance, the indirect measure may be a priming effect in response times generated by a *critical feature* of the prime (e.g., its color), and the direct measure may be some measure of awareness of that critical feature (e.g., objective discrimination performance or subjective visibility ratings). To appreciate the full range of possibilities of the dissociation paradigm, consider Figure 1a. If we plot performance in the indirect measure against performance in the direct measure, we span a *D-I space* (Schmidt, 2000; Schmidt & Vorberg, 2006).

There is a horizontal line where the indirect effect is zero and a vertical line where the direct effect is zero. If we express *D* and *I* in the same metric (Schmidt & Vorberg, 2006; e.g., the effect-size metric *d'* of Signal Detection Theory [SDT]), the main diagonal indicates cases where direct and indirect effects are identical in magnitude. For cases falling above the diagonal, the indirect effect is larger than the direct one, and vice versa below the diagonal. Different types of dissociations appear as distinct patterns when plotted in *D-I* space. We are especially interested in data patterns where each of the two measures appears as a function of a parametric variable, e.g., a variation in timing or contrast. When the parametric course of the two functions lead to interesting discrepancies, we speak of *functional dissociations*. They appear as *dissociation curves* in *D-I* space.

Figure 1 show various possibilities of dissociation curves between direct and indirect measures (Schmidt & Vorberg, 2006; Schmidt & Biafora, 2024). *Simple dissociations* (Fig. 1b) occur when the critical feature induces an indirect effect while a direct measure of awareness for that feature shows zero performance. Examples for this data pattern can be seen in Vorberg et al. (2003) and F. Schmidt & Schmidt (2010). Note that the reverse pattern (variation in awareness without any indirect effect) also constitutes a simple dissociation and can also be of theoretical interest.

Simple dissociations are a special case of *invariance dissociations*, which are not confined to null effects (Fig. 1c). Generally, variation in *I* at an invariant level of *D* (circles) indicates that changes in the direct measure are not *necessary* for changes in the indirect measure, while variation in *D* at an invariant level of *I* (squares) shows that changes in the direct measure are not *sufficient* for producing changes in the indirect measure. The Fehrer-Raab effect (Fehrer & Biederman, 1962; Fehrer & Raab, 1962) is a classic example of an invariance dissociation where response times to a masked stimulus remain constant under variations in awareness (metacontrast masking). Similarly, Lamy et al. (2017) argue that response priming remains unaffected when visual awareness for a masked prime is enhanced by previous presentation.

Meyen et al. (2022; Zerweck et al., 2021) argue that dissociations are only convincing if the indirect effect is clearly larger than the direct one. Figure 1d shows an example of such a *sensitivity dissociation* (Reingold & Merikle, 1988). Sensitivity dissociations (but not the other dissociation types) require equal scaling of the two measures as well as a measurement-theoretical assumption that the direct measure responds at least as sensitively to changes in the critical feature as the indirect measure (Schmidt & Vorberg, 2006). [1]

[1] Several techniques have been proposed for expressing direct and indirect effects on the same scale. Response time effects are often converted into *d'* units by performing a median split on the distribution and then cross-tabulating fast and slow responses with prime-target congruency to generate hits and false alarms (Schmidt, 2002; Meyen et al., 2022; Zerweck et al., 2021). However, this down-transformation of RT to a dichotomous measure loses valuable information. On the basis of simulations, Schmidt and Vorberg (2006) recommend calculating *d'* directly from the distance between the RT distributions, after normalization by the pooled standard deviation and Winsorization of outliers.

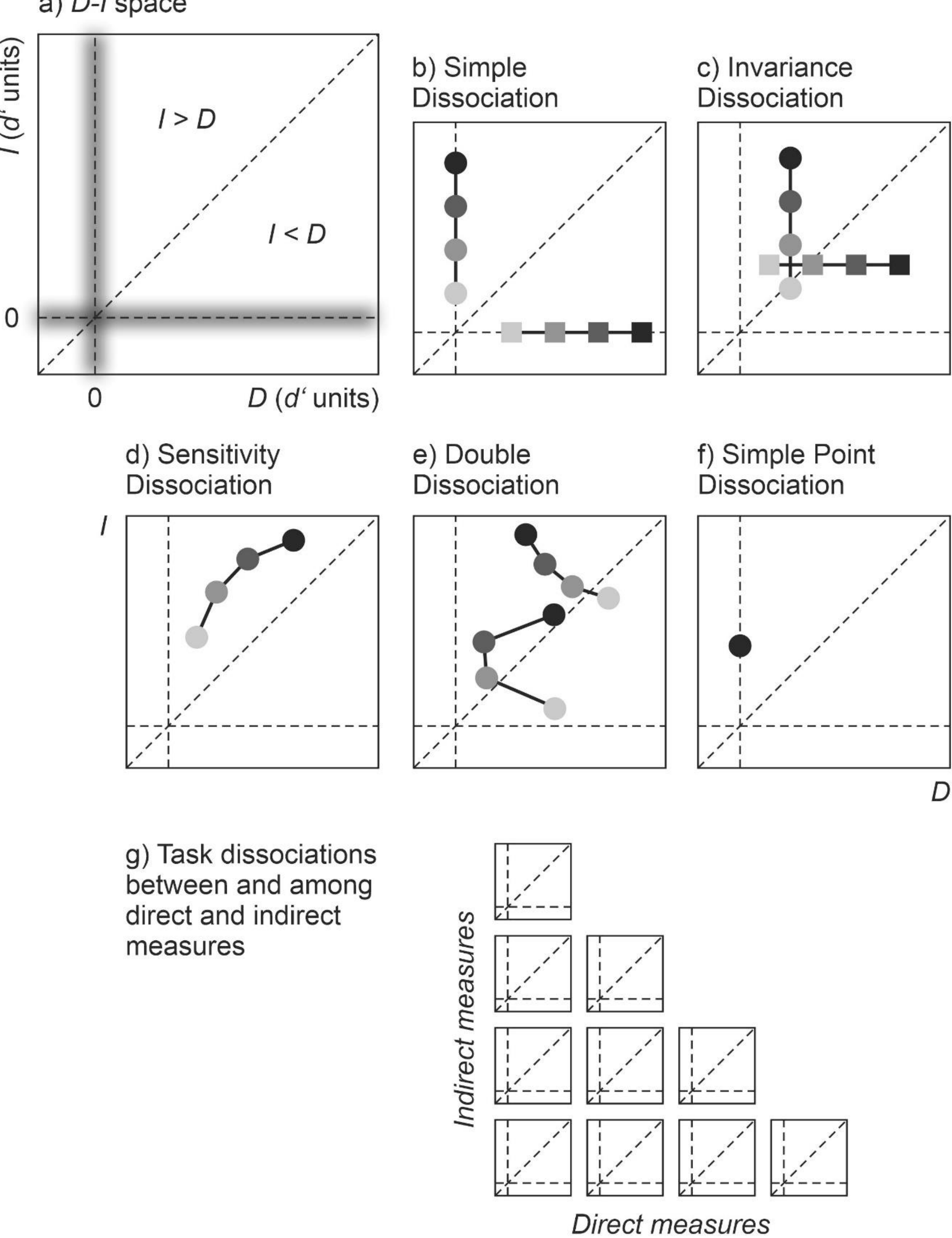

*Figure 1: Dissociation curves between a direct and an indirect measure. a) In a* D-I *space, indirect effects are plotted against direct effects after conversion to a shared metric (here,* d'*). Fuzzy lines at effect sizes of* D = 0 *and* I = 0 *symbolize the statistical problem of determining whether effects differ from null performance. b) Simple dissociations show variation in one effect while the other effect is zero. c) Invariance dissociations describe the more general case where the invariant effect is constant at some value other than zero. d) Sensitivity dissociations show one effect to be clearly larger than the other. e) Double dissociations show that an experimental manipulation has opposite effects on both measures: The two examples show increasing indirect*

*effects under either decreasing or u-shaped direct effects (e.g., "type-B masking"). f) Of all those possibilities, Stockart et al.'s paper only discusses a special case of b), the simple point dissociation. g) Task dissociations can occur not only between but also among direct and indirect measures, each of which is based on its own decision space and criterion content (Kahneman, 1968). – In all plots, the shading of data points indicates some experimental variation that leads to increasing indirect effects, like increasing prime-target interval or prime intensity; however, dissociation patterns are defined solely by their shape in* D-I *space, not their directionality.*

Simple and invariance dissociations share a fundamental problem: We cannot infer invariance of awareness from invariance of the direct measure unless we can assume that this measure is both *exhaustively valid* (Schmidt & Biafora, 2024) and *exhaustively reliable* (Reingold & Merikle, 1988). Exhaustive validity means that the direct measure must capture all information sources theoretically relevant for the direct task. In particular, the direct measure is required to capture awareness of the *critical feature* (the feature that drives the indirect effect; e.g., prime color in a color priming task). Exhaustive reliability means that the direct measure must be able to pick up any variation in those relevant sources of information, no matter how small – an assumption that is very problematic from a psychometric viewpoint (see Schmidt & Biafora, 2024, for a formal definition of exhaustiveness). Fortunately, there is a data pattern that does not require exhaustive reliability at all: *double dissociation*. A double dissociation is demonstrated when the two measures show opposite courses under parametric variation of an independent variable, for instance, increasing priming effects despite decreasing performance in prime discrimination. When plotted in D-I space, double dissociation leads to a dissociation curve that slopes downward somewhere along its course (Fig. 1e). A double dissociation immediately implies that the direct measure cannot explain variation in the indirect one, and it only makes minimal assumptions about the underlying measurement theory (Schmidt & Vorberg, 2006). In research on unconscious perception, double dissociations have primarily been observed for response priming under metacontrast masking. Under suitable stimulus conditions, metacontrast can lead to masking functions that are decreasing or u-shaped when prime-mask SOA is varied, whereas priming effects increase with SOA (Vorberg et al., 2003; Biafora & Schmidt, 2020, 2022; Mattler, 2003; cf. Dunn & Kirsner, 1988; Merikle & Joordens, 1997, for closely related concepts). Double dissociations can be hard to find, but recently we have developed an experimental technique that can help bend masking functions into a desired shape without affecting response activation from the prime (Biafora & Schmidt, 2020). This technique of "induced dissociations" can be transferred to domains other than metacontrast masking.

The overwhelming majority of studies in unconscious cognition is based on simple dissociations with their requirement for exhaustive reliability. Exhaustive reliability is a rather severe assumption because any type of measurement noise in the data (and there is usually plenty in binomial forced-choice and multinomial rating data) will violate the assumption and call the conclusion of unconscious processing into question (Malejka et al., 2021; Reingold & Merikle, 1988). But apart from this measurement-theoretical problem, a simple dissociation is very difficult to establish to begin with because no procedure exists that ensures chance-level masking at the group level of observers. Even though there are lawful relationships between stimulus variables and the average shapes of masking functions (Breitmeyer & Öğmen, 2006;

Kahneman, 1968), masking functions still vary quantitatively and qualitatively from person to person. Even with the same set of stimuli, individual observers' performance may be at chance, or at ceiling, or show increasing, decreasing, or u-shaped masking functions. As fascinating as these individual variations often are, they are problematic when hunting for chance-level effects at the group level: For the group-averaged masking function to be close to chance, the great majority of individual observers must be close to chance in the first place, and that simply does not occur so easily. Staircase or threshold procedures may sometimes help with idiosyncratic masking functions, but more often than not fail to compensate for the qualitative differences in data patterns. This has a very important and underestimated consequence for the entire classical dissociation logic: simple dissociations depend on sheer luck. Only once in a while we have the good fortune of finding a group-averaged masking function close to zero with small error bars and good agreement between observers, and whenever that happens we have a "proof of concept" that, say, response priming can occur without prime discrimination (Vorberg et al., 2003, Exp. 1, or F. Schmidt & Schmidt, 2010). But as soon as we repeat the experiment, we will have a new group of observers with their idiosyncratic masking functions. Fortunately, we can measure each individual observer with high precision by using massively repeated measures, so our problem is actually not measurement noise (failure to replicate at the individual level) but real idiosyncrasy (high replicability within observers, but variation of data patterns between observers). This dependence of the simple dissociation paradigm on blind luck has certainly been a major driving force in developing methods that promise to identify "unaware" trials irrespective of the concrete shape of the masking function. [2]

Stockart et al. do not seem to be aware of the possibilities of functional dissociation patterns. They discuss their methods by means of a simple priming paradigm (based on Dehaene et al., 1998) that generates only a single data point on the single line in *D-I* space where *D* is exactly zero. We call this scenario a *simple point dissociation* (Fig. 1f); it is a special case of a simple, invariance, or sensitivity dissociation crunched into a single coordinate. Many of the problems with Stockart et al.'s recommendations arise from the attempt to draw spurious information from this single data point – specifically, from the noise in the data point.

We now look at what we consider the most important of Stockart et al.'s recommendations. We focus on the list presented in their Figure 2, using the same numbering but our own thematic sorting (some points appear more than once, some not at all). We urge readers to read their text in detail to appreciate the full range of pro and contra arguments the authors provide. Note that most of their recommendations are based on a solid majority of votes, and only a few recommendations met with some dissent.

[2] There are several constructive ways to deal with idiosyncratic masking functions. In metacontrast masking, individual masking functions are stable over time, and observers can often be classified into having increasing (type-A) or u-shaped (type-B) masking functions (Albrecht et al., 2010; Albrecht & Mattler, 2010, 2012, 2016). So, one avenue would be extensive pretesting under a set of predetermined criteria, followed by contrasting groups of pretested and re-validated observers to demonstrate increasing priming for masking functions of both directions. A second strategy is to take the individual level more seriously: In any group of well-measured observers, there is usually plenty of evidence that the time-course of masking does not predict the time-course of priming. It is straightforward to integrate such evidence statistically across single observers.

## 2. Choice of measures and measure-specific interpretation

*"A3: Subjective awareness measure: A preference towards the PAS."*

*"A4: Objective direct measure: Use a forced choice discrimination task on the feature of interest."*

*"C9: Precisely define what is meant by 'unconscious processing'."*

*"C10. Justify the chosen measures and provide conditional interpretations."*

Any dissociation between one direct and one indirect measure is only a special case of a larger space of *task dissociations* (Fig. 1g). This approach acknowledges that there are not only dissociations *between* direct and indirect measures, but also *among* them (Koster et al., 2020; Schmidt & Biafora, 2024). Direct and indirect tasks differ in the *decision spaces* they require, which predict distinct patterns of sensitivity and response biases, e.g., when detection, yes-no discrimination, and same-different tasks are compared (Hautus et al., 2022). They also differ in the *criterion content* they are targeting, e.g., whether they are asking for the color, shape, or lexicality of a stimulus (Kahneman, 1968). Criterion content can be characterized as the *set of cues* individual observers use to perform a task: e.g., when trying to discriminate a masked prime, they may use cues from visual motion, flicker, perceived response conflict, or task knowledge in addition to their actual awareness of the critical feature (*Cue Set Theory*, Schmidt & Biafora, 2024).

When measuring the visibility of a stimulus, we have to deal with the fact that visual awareness consists of a multitude of facets that must be measured separately and may even undergo double dissociations among each other (Koster et al., 2020). Even for a single facet of visual awareness, multiple methods exist to measure it, both subjectively and objectively, and each comes with its own decision space and criterion content. In visual masking alone, commonplace methods contain stimulus rating, stimulus matching, direct scaling, multidimensional scaling, and of course the whole variety of detection and discrimination tasks (Breitmeyer, 1984; Breitmeyer & Oğmen, 2006; Hautus et al., 2022; Sackur, 2013). Therefore, various dissociations among direct tasks are to be expected, are of theoretical interest, and should be studied extensively. In fact, dissociations between different direct tasks are abundant in the literature on visual awareness. Different tasks have been used to address specific criterion contents, asking for position, shape, motion direction, lexicality, object class, and countless other stimulus properties. In general, masking functions are not identical across tasks. For instance, metacontrast masking largely spares the ability to detect a target but strongly affects brightness matching and contour discrimination (Breitmeyer & Oğmen, 2006), and targets that are easy to detect can be very difficult to discriminate (Vorberg et al., 2003). Dissociations between indirect tasks, on the other hand, are just as common – for instance, responses made with different effector systems, different numbers of response alternatives, or different speed-accuracy instructions.

We therefore agree with Stockart et al. that different types of direct measures should be considered (e.g., objective and subjective ones), that they should be chosen carefully, and that interpretations and inferences should be specific to the measures employed. But in the light of the extraordinarily rich literature on task dissociations we find it puzzling that Stockart et al. restrict their recommendations to only two measures, the Perceptual Awareness Scale (*PAS*; Ramsøy & Overgaard, 2004) and forced-choice tasks (obviously, that includes yes-no discrimination [YN] and two-alternative forced choice [2AFC]; Hautus et al., 2022). We have previously criticized the PAS on several grounds (Schmidt & Biafora, 2024). Firstly, the original wording of the PAS focuses mainly on detection instead of discrimination of the critical stimulus (and also a bit on decision confidence), and will therefore typically fail to address the critical feature that generates the indirect effect. Secondly, when Ramsøy and Overgaard (2004) introduced the PAS, they were careful to apply the ratings to different stimulus dimensions separately (i.e., there were different PAS ratings for color, shape, and position). Without such separation of stimulus aspects, the PAS only focuses on the general visibility of "the stimulus" as a whole, not any particular well-defined facet of awareness. Thirdly, we criticize the common practice of changing the wording of the scale labels from experiment to experiment (expressly encouraged by Sandberg & Overgaard, 2015), which basically obliterates the PAS as a well-defined psychometric tool. Today, "PAS" seems to be an umbrella term for a loose variety of custom-made rating scales that are sometimes not even explicitly described. From a psychometric perspective, any adjustment of the PAS to a novel criterion content creates a new scale with new measurement properties (including its reliability and validity structure).

In contrast, we fully agree with Stockart et al.'s recommendation that interpretations must be specific for the measures employed. Instead of loosely claiming "unconscious perception" or "subliminal priming", we should state clearly which measures are dissociated, and in what way (for instance, "increasing effects of color priming despite invariant prime discrimination"). Importantly, we should be prepared for the discovery that the particular dissociation we observe does not readily translate to other direct or indirect measures (Koster et al., 2020). Because different facets of awareness are interesting in their own right, they should not be viewed as in need of "justification" as "measures of consciousness". Rather, they should be motivated based on the criterion content they are addressing, e.g., detection, color discrimination, lexical decision, and so on. The same applies to the indirect measure.

## 3. Measurement precision and multitasks

*"A1: Ensure adequate precision of the processing and awareness measures."*

*"A2: Use both subjective and objective direct measures."*

*"A5: Collect all measures on a trial-by-trial basis."*

A2 and A5 together lead to a recommendation for dual or even triple tasks. We will argue that they clash with A1's demand for adequate measurement precision.

First of all, we naturally agree with recommendation A1 that all measurements should be high-powered and precise. We are strongly opposed to the common practice of employing weak, underpowered measures for the direct task. Examples would include restriction to only a short "post-test" [3], the use of weak criteria for unawareness (e.g., nonsignificant correlations between direct and indirect measures calculated across a small number of participants), or linear extrapolation from clearly discriminable to purportedly unseen stimuli (see Stein et al., 2024, for a forceful critique of such practices). In our own research, our goal is to have enough trials per data point and participant to evaluate direct and indirect effects in single observers, not just as group averages (F. Schmidt, Haberkamp, & Schmidt, 2011). This is why we have a strong preference for designs with a relatively small array of observers but high measurement precision within observers ("small-*N* / large $N_i$" designs"; Smith & Little, 2018; see Arend & Schäfer, 2019; Baker et al., 2021, for proof that such designs can have excellent statistical power even at the group level). The level of individual participants is especially important when visual masking is involved, because masking effects vary widely and qualitatively across observers, to a degree that averaging them can be downright misleading (Albrecht et al., 2010; Albrecht & Mattler, 2010, 2012, 2016; Biafora & Schmidt, 2022). Many studies in visual masking therefore take care to show the results of individual participants, in line with the research tradition of psychophysics. Of course, we do acknowledge that there are some paradigms and methods that require larger numbers of observers, especially when based on correlation between measures (Vadillo et al., 2024).

As already stated, we concur with recommendation A2 that subjective as well as objective measures should be used – but potentially all of them, not only PAS and forced-choice discrimination. For example, Koster et al. (2020) first asked participants to describe their subjective percepts of metacontrast-masked shape primes. Next, the reported percepts were sorted, classified, and converted into rating scales that were applied as direct tasks in a second part of the study. It turned out that different percepts had qualitatively different time-courses, some increasing with prime-mask

[3] In the polls on which the recommendations are based, the collective was faced with an improper alternative regarding the question "When should the objective test be administered?". The alternatives were named "Trial-by-trial", "During the main task, in separate trials", and "Post-test". While this wording is unclear enough, the obvious fourth alternative of taking the direct measure in entire separate *sessions* (i.e., on equal footing with the indirect task) to measure a complete masking function was never offered.

SOA, some decreasing, and some increasing only when prime and target were consistent in shape but not when they were inconsistent. Objective discrimination performance always increased with SOA and was therefore double-dissociated from several of the other awareness measures. This seminal study clearly demonstrates that different measures can be based on widely different criterion contents and that no single measure will be able to capture the contents of subjective awareness as a whole (Schmidt & Biafora, 2024). This is true for both measures advocated by Stockart et al., forced-choice discrimination and PAS ratings.

To us, the third recommendation (A5) seems most problematic. Collecting direct and indirect measures "on a trial-by-trial basis" is supposed to mean "on the same trial", i.e., in a dual or even triple task. For instance, Peremen and Lamy (2014) asked participants to perform three tasks on each trial of several masked-priming experiments: first a speeded response to the pointing direction of the target (2YN), then an unspeeded response to the direction of the prime (2YN), and finally a PAS rating (on an apparently modified scale not fully described). Biafora and Schmidt (2022, Exp. 1) compared such a triple task with the corresponding single tasks. Their experiment had six sessions. The first three were entirely devoted to the target response, the prime response, and the PAS rating (in its original wording), respectively; in the final three sessions all three tasks were performed on each trial in the same sequence as in Peremen and Lamy (2014). We found that the triple task completely changed the pattern of results in all three measures involved. Compared to single tasks, the triple task yielded much higher prime discrimination performance, but without a correspondent increase in PAS ratings. More importantly, it drastically altered the temporal structure of the priming effects. The triple task led to a striking 160-ms delay in response time to the target, a marked reduction in fast errors, and a reduction in the time-locking of errors to the prime. Note that these changes all indicate that response priming was no longer characterized by fast feedforward processing. Even more importantly, there were qualitative changes in data patterns. While both single and triple tasks showed decreasing (type-B) masking functions in both direct measures, priming effects increased in the single task but *decreased* in the triple task – a data pattern extremely unusual in response priming. In other words, a highly informative double dissociation between priming and masking (stronger priming in spite of decreasing discrimination) turned into an uninformative positive association. Kiefer et al. (2023) report similar problems in the domain of semantic priming. We concluded that the multiple responses per trial had to be collected and maintained in working memory before being reproduced in sequence, and that this operation simply alters (and damages) the nature of processing for all variables involved. We analyzed the data at the group level, but also observer by observer, and found many surprising dissociations between forced-choice performance and PAS ratings: whereas prime discrimination performance was generally higher in the multitask, PAS ratings were higher in some observers but lower in others.

Even though the higher discrimination accuracy may have been a result of differential practice (the triple task was always performed in later sessions than the single task because we expected it to be harder), there is always a danger that reconciling the information from all the concurrent tasks may provide additional cues to the identity of the prime. For example, if participants developed the (correct) hypothesis that inconsistent primes provoke response errors, a successful strategy for improving prime discrimination in multitasks would be to choose the response opposite to the target on each error trial. Relatedly, Li et al. (2024) have shown that concurrent confidence ratings can reactively improve accuracy but prolong response times, which they explain by more conservative boundary settings in a drift-diffusion model.

Our conclusion from our multitask studies was that triple tasks (and possibly even dual tasks) are not well suited for measuring the relations between priming and masking because they require a massive cognitive load that impairs all measurements (Biafora & Schmidt, 2022). It is, however, theoretically interesting to see *how* and *why* those measurements are impaired. This question is especially interesting in the light of an argument by Peremen and Lamy (2014), who attempt to turn our own reasoning upside down. They suggest that the double dissociations observed between increasing response priming and decreasing masking functions (Vorberg et al., 2003) are actually an artifact of attending to the target in one task but to the prime in the other. In that view, only multitask measurements allow for an unbiased comparison between priming and masking because they somehow equalize those attentional differences. Apart from the fact that this argument calls the entire literature on visual masking and response conflict paradigms into question, which has invariably employed single tasks, it is not clear to us why the equalization of attention should work in the first place. It seems more likely that attention is selectively withdrawn from the target because the prime-related tasks are simply more cognitively demanding. In any event, the massive time delay introduced by the multitask situation presents a problem for both response priming and masking. In priming, slower responses are associated with response inhibition (*negative compatibility effect*; Eimer & Schlaghecken, 1998) as shown by analyses of response time distributions (Panis & Schmidt, 2016; Schmidt et al., 2015). And in metacontrast masking, masking functions can change from increasing (*type-A*) to u-shaped (*type-B*) when observers are required to respond more slowly (Lachter & Durgin, 1999). These findings suggest, unsurprisingly, that it is rather the triple task that tends to invite artifacts. For these reasons, the recommendation for multitasks not only clashes with the requirement of "adequate precision" (A1), but also of measurement validity.

To be able to evaluate performance in the first place, direct and indirect tasks should be explored for their own sake – and on their own terms. This means that researchers interested in unconscious processing need to pay attention to quality standards in adjacent fields like visual masking, visual rivalry, semantic priming, or the continuum of response priming, flanker, and inhibition effects (Ansorge et al., 2014; Breitmeyer & Öğmen, 2006; Ellinghaus et al., 2024; Panis et al., 2016, 2020; Sterzer et al., 2014). For instance, many experiments have analyzed the time course of response activation and inhibition in response priming (e.g., Lingnau & Vorberg, 2005; Schmidt et al., 2022; Schmidt & F. Schmidt, 2009, 2010; Vorberg et al., 2003; Wolkersdorfer et al., 2020). All of these experiments were carried out as single tasks, and many of them employed unmasked primes to explore the properties of response priming without the additional complication of masking effects (e.g., Haberkamp et al., 2013; F. Schmidt & Schmidt, 2014; Seydell-Greenwald & Schmidt, 2012). These studies suggest that response priming is based on strictly sequential waves of response activation elicited in turn by prime and target, thus forming a very simple system of sequential feedforward activation. Understanding the feedforward nature of unmasked response priming then helped explain the dissociation between priming and masking: Because research on backward masking had already led to evidence that it is a time-delayed process based on recurrent or reentrant signals (Breitmeyer & Öğmen, 2006; Breitmeyer et al., 2004; Di Lollo et al., 2000; Fahrenfort et al., 2007; Francis, 2000; Kammer, 2007; Lamme et al., 2002; Pascual-Leone & Walsh, 2001; also see Becker & Mattler, 2019), it was plausible that an initial, prime-triggered feedforward wave of motor activation might simply escape the slower masking process (*Rapid-Chase Theory*, Schmidt, Niehaus, & Nagel, 2006). This example of theoretical integration across fields illustrates that in order to understand masked priming, we not only have to understand masking, but priming as well. To us, that implies that the methods we employ in masked priming should not be too disparate from those employed in adjacent fields. If we employ all measures of priming and masking in dual or even triple tasks, we deviate strongly from the standards under which these variables are usually measured and can no longer rely on the knowledge that has been accumulated about those tasks.

### 4. The post-hoc selection fallacy

*"A5: Collect all measures on a trial-by-trial basis."*

*"B8: Examine the possible effect of misclassification due to measurement error."*

It may not be apparent at first glance that these recommendations deal with the strongly criticized practice of post-hoc selection of "unconscious" trials on the basis of visibility ratings, or the selection of "unconscious" observers on the basis of prime discrimination performance. Taken singly, both recommendations reached a strong consent. There was, however, no vote on the practice of post-hoc selection itself.

Stockart et al. extensively discuss the well-known problems of *regression to the mean (RttM)* arising in post-hoc selection (Shanks, 2017; also see Malejka et al., 2021; Rothkirch & Hesselmann, 2017; Rothkirch et al., 2022; Shanks, Malejka, & Vadillo, 2021; Vadillo et al., 2022). When trials from the lowest rating category are selected, unreliability of the scale causes some of those trials to land in the lowest category only because of measurement noise. As a result, the visibility of those trials is underestimated (conversely, the visibility in the highest rating category would be overestimated). This phenomenon is sometimes called a *squeeze effect*: no matter whether scores are extremely high or extremely low, they tend to be more moderate upon replication. Characteristically, squeeze effects also occur retrospectively: extreme scores in the replication turn out to have been more moderate in the first run. Regression to the mean is an inevitable consequence of any less-than-perfect reliability in the direct measure, and the lower the reliability, the stronger the squeeze (Campbell & Kenny, 1999).

Even though Stockart et al. propose some mathematical correction techniques for regression to the mean, they do not discuss the consequences for the simple dissociation paradigm they advocate. First of all, the techniques proposed so far usually have to make relatively strong assumptions to accomplish the correction. For instance, the correction recently proposed by Dienes (2024) assumes that there are only two perceptual states (conscious or unconscious), that there is no response bias in their subjective classification, and that misclassification rates follow a particular Bayesian prior function (in contrast, see Yaron et al., 2024, for a nonparametric approach). Second and more important, those corrections move the estimate away from the point of zero awareness. Corrections for regression to the mean therefore make it even harder to find a simple dissociation: Not only do we have to be lucky in producing an initial estimate that falls as closely as possible to the zero-visibility line, it also has to remain sufficiently small after correction for RttM. Stockart et al. appear quite optimistic about the possibility of correcting for the regression artifact, but fail to mention that the correction will likely destroy the simple dissociation.

But regression to the mean, as annoying as it is, is not the greatest problem here. In an earlier critique of post-hoc sorting, the practice was characterized as a "sampling fallacy" and as "capitalizing on chance" (Schmidt, 2015). Stockart et al.'s suggestion of collecting both a forced-choice response and a PAS rating on the same trial allows for a striking illustration of this fallacy if we realize how those two measures are psychophysically related. Consider the familiar signal-detection model of a simple yes-no discrimination task where one of two stimuli is presented on each trial and the participant has to decide under uncertainty which one it is. The standard SDT model postulates that the internal evidence for one versus the other stimulus alternative is measured by two overlapping distributions, *S1* and *S2* (Fig. 2A). Variation within these distributions is assumed to arise solely from random noise. The two distributions vary along the horizontal *decision axis* that ranges from evidence in favor of *S1* to evidence in favor of *S2*, with a point of uncertainty in between. For the objective discrimination task, a criterion is placed on the decision axis, and responses for *S1* or *S2* are chosen depending on which side of the criterion the evidence falls. For a signal-detection analysis, we can arbitrarily define *hits* when Stimulus 2 is correctly identified as "Stimulus 2", and *false alarms* when Stimulus 1 is incorrectly identified as "Stimulus 2".

Why low visibility ratings do not imply low sensitivity

a) How rating criteria subdivide the decision space for any given value of *d‘*:

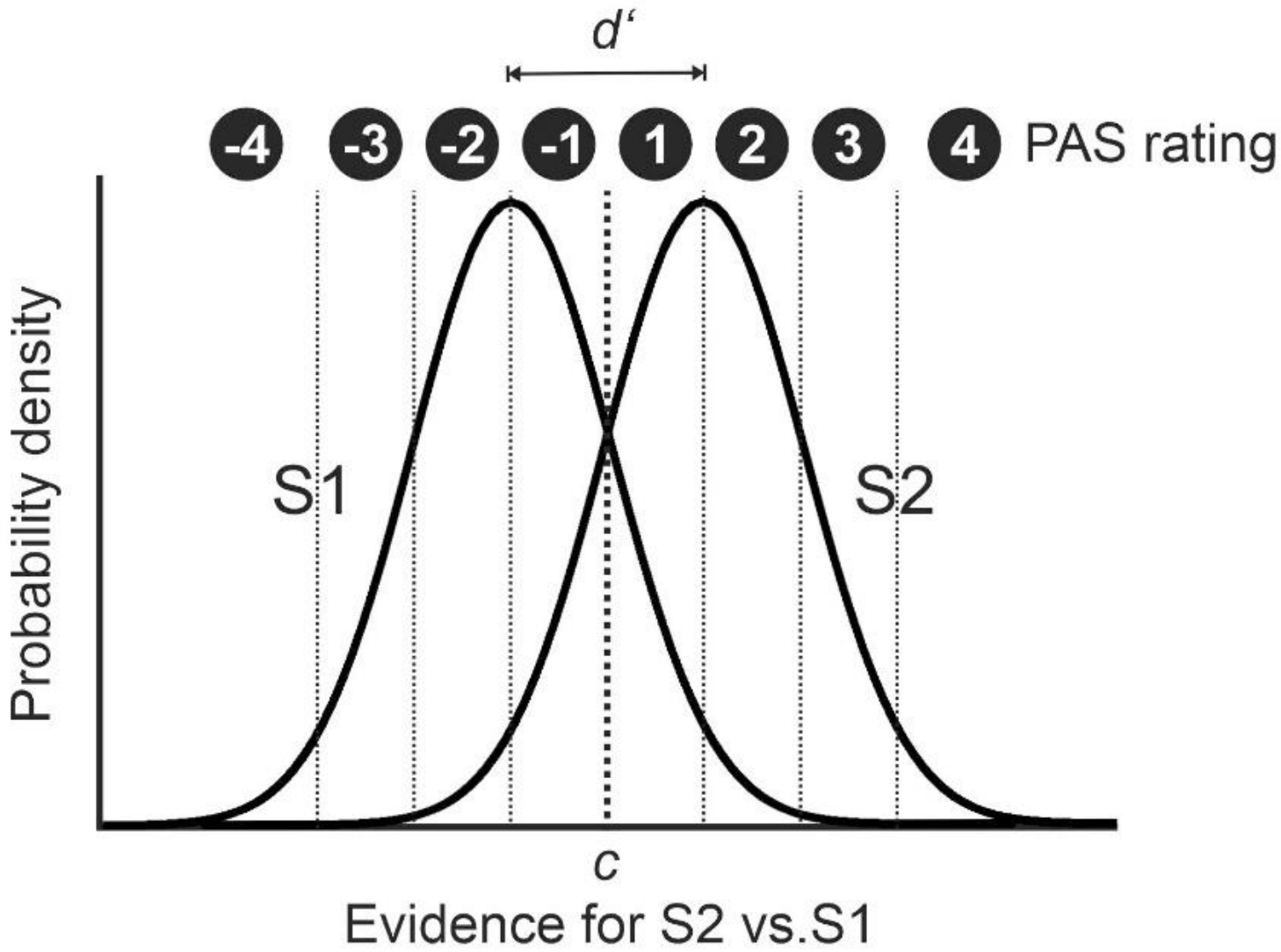

b) Any ROC curve must pass the point of indifference between stimulus alternatives:

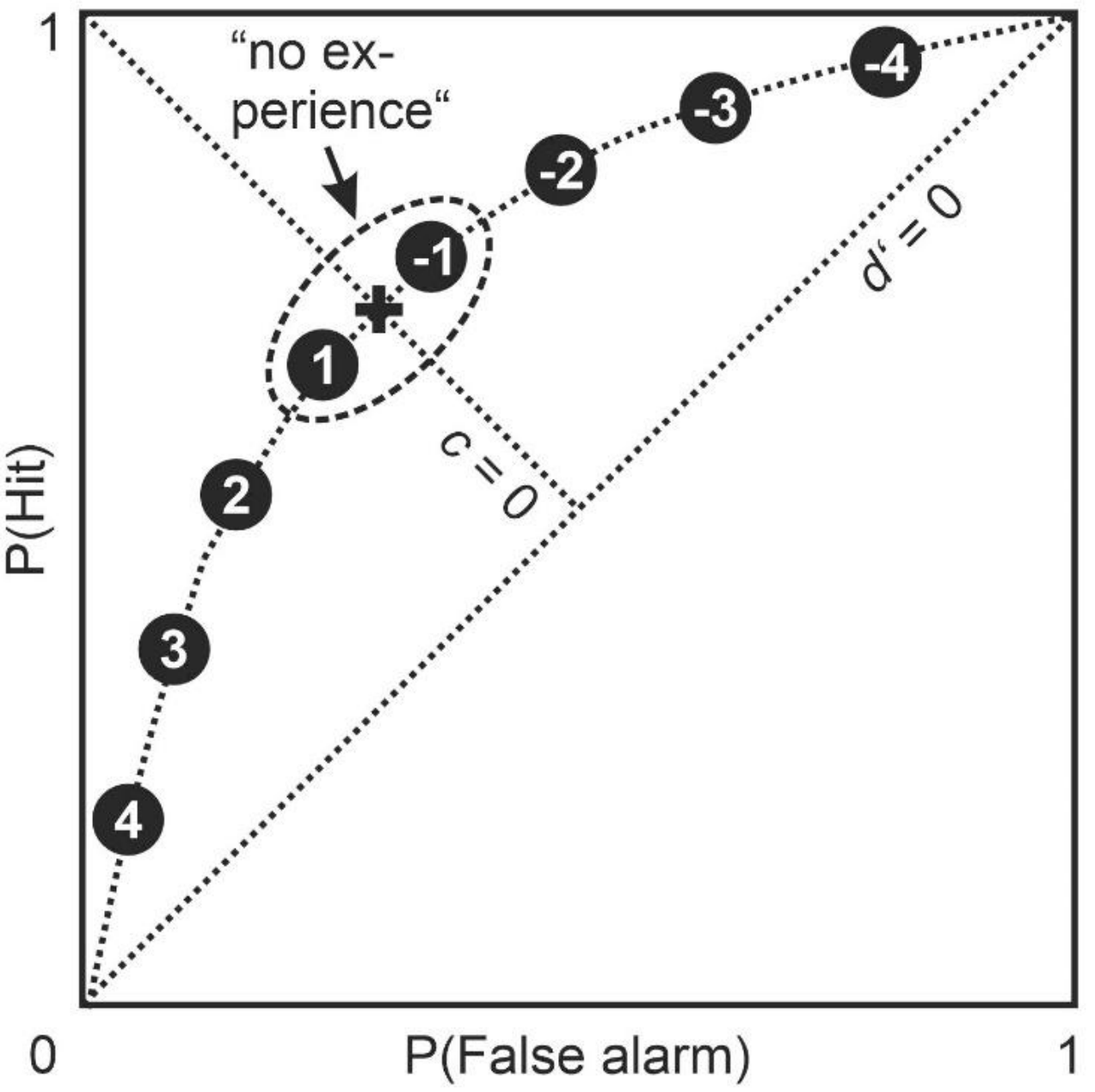

*Figure 2: Signal detection model for joint PAS ratings and objective discrimination judgments in a single experimental condition. A) Internal evidence for Stimulus 1 versus Stimulus 2. For simplicity, both stimuli are assumed to lead to identical standard normal distributions along the “S2 vs. S1” decision axis, so that the sensitivity index* d′ *simply corresponds to the distance between the distributions. The objective discrimination criterion (*c*, heavy vertical line) is assumed to be optimal here, minimizing the number of classification errors. Hits are arbitrarily defined as “S2” decisions to S2 stimuli, false alarms as “S2” decisions to S1 stimuli. PAS visibility ratings range from 1 to 4, with a negative sign attached whenever the critical stimulus is identified as S1. Rating categories are separated by additional response criteria (light vertical lines). B) The ROC curve plots cumulative rates for hits versus false alarms for PAS categories in ascending order. Major diagonal: marks* d′

*= 0 where hit rate equals false alarm rate. Minor diagonal: marks neutral response criterion* c *= 0. Dotted curve: marks an isosensitivity curve where all points have the same sensitivity* d′ *but may differ in response criterion. The plus sign marks the spot on the ROC where S1 and S2 are theoretically indistinguishable to an ideal observer. The lowest ratings (±1) are marked with the PAS label "no experience" (Ramsøy & Overgaard, 2004). Note that they are part of the same ROC as all the other rating categories and thus indicate the same sensitivity.*

There is a standard way for representing visibility ratings within this decision space. Stockart et al. correctly recommend that both the PAS labels and the forced-choice task should be designed to reflect the critical feature that generates the indirect effect (as originally intended by Ramsøy & Overgaard, 2004). This helps to bring the decision axes for PAS ratings and forced-choice discrimination into register so that both tasks ideally use the same axis. This leads to the SDT standard model for a single condition in a rating experiment where the different rating categories will cut the plot into vertical stripes (Fig. 2A; Gescheider, 1997; Green & Swets, 1966; Hautus, Macmillan, & Creelman, 2022). The lowest ratings are placed in the center of the plot where the overlap between the distributions is largest; the next highest ratings come from the two stripes left and right to the center category; and so on towards the tails of the distributions. The different PAS categories are separated by their own (subjective) criteria that are used to determine the category in which a rating response is placed. Those categories can be used to plot the cumulative hit rate against the cumulative false alarm rate in a *receiver operating characteristic (ROC)* curve. The curve starts in the lower left corner with the high-visibility ratings for Stimulus 2, continues through progressively lower ratings for Stimulus 2, switches to low ratings for Stimulus 1, and ends on the upper right with the high-visibility ratings for Stimulus 1 (Fig. 2B). Note that if the standard SDT model applies (normal distributions, equal variances), we do not even need low visibility ratings from any actual observer: from our knowledge of *d′* alone the entire ROC is determined, including the point where it crosses the minor diagonal of the plot (marked with an "x" in Fig. 2B). This is the theoretical point where an optimal observer would have no idea whether S1 or S2 were presented.

It may be surprising to many users of post-hoc trial sorting that in this standard model the PAS ratings are *unrelated* to perceptual sensitivity. This is because sensitivity and response bias are defined on the level of the *distribution* of trials, whereas PAS ratings are defined on *single* trials. As soon as the sensitivity is given and the shape and position of the ROC is known, the different positions along the curve only arise from the noise in the decision, and the different rating categories represent different decision criteria along this continuum. For a discrimination task, the lowest PAS ratings lie right in the center of the ROC where the curve crosses the minor diagonal (the theoretical point where response bias is zero). This means that low PAS ratings can only be interpreted when the underlying sensitivity is known. In addition to measuring the visibility on every single trial, it is therefore necessary to consider the *overall* sensitivity from the entire *distribution* of trials in any experimental condition.[4]

---

[4] Proponents of subjective criteria of awareness may complain that we insist on an "objective" sensitivity standard as a precondition for subjective ratings, but that would be a misunderstanding. SDT is agnostic about any concepts of consciousness or subjectiveness, and the decision axis in Fig. 2A may well be conceived as subjective information for S1 versus S2. The critical question is not whether an objective measure allows for the interpretation of a subjective one, but whether the information *on the*

This analysis makes clear in which sense the isolated analysis of low PAS ratings is a sampling fallacy: it constitutes a negative selection from an underlying distribution that could be consistent with different levels of sensitivity – and that may or may not be consistent with the hypothesis of unconscious processing. In our own data (Biafora & Schmidt, 2022, Exp. 1), PAS ratings of 1, 2, 3, and 4 corresponded to prime discrimination accuracy $p_c$ of .590, .643, .722, and .933, respectively (averaged across all observers and conditions). Post-hoc sorting by PAS ratings would therefore conclude that discrimination accuracy in the lowest rating category was at 59.0 % (which would correspond to $d' \approx .455$, using the approximation that $d' \approx z(p_c) - z(1\text{-}p_c)$; Hautus et al., 2022). Even if correct, this value would be too high to meaningfully speak of unconscious stimulus conditions, but it is in fact a gross underestimation of the true underlying accuracy across all ratings, which is 68.8 % ($d' \approx .980$). This simple example makes clear that low PAS ratings cannot be interpreted in isolation (Schmidt, 2015). They may successfully isolate trials with low subjective *visibility*, but those will necessarily occur at any level of sensitivity – after all, every discrimination ROC has to cross the minor diagonal! Therefore, low PAS ratings do not isolate trials with low *sensitivity*, which remains unknown unless the entire distribution of trials is considered.

Regression to the mean contributes to this problem, but probably not in the way users of post-hoc selection would expect. The artifact occurs because an imperfect reliability in the PAS ratings causes some single trials to be misplaced into extreme rating categories. Upon retesting, the regression artifact would lead them to be squeezed back into categories closer to the center of the trial distribution (Fig. 3, left panel). Importantly, regression to the mean in the PAS ratings does not change the underlying ROC, it only leads to misclassification *along* the curve: it affects only trial-by-trial visibility, but not condition-wide sensitivity. It is important to contrast this effect with the other widespread practice of discarding entire participants on the basis of their objective discrimination performance. Selectively analyzing low-performing observers constitutes a negative selection of observers, and once again unreliability in the measurement leads to regression to the mean. But this time, it works in the direction of sensitivity, not visibility (Fig. 3, right panel): by selecting only low-performing observers, we underestimate the performance they would show upon retesting.

---

*level of the distribution* allows for the interpretation of information *on the level of single trials*. Objective sensitivity as measured by $d'$ is only one example of a distribution-wide measure, but any distribution-wide subjective measure (e.g., average visibility ratings) would create the very same problem.

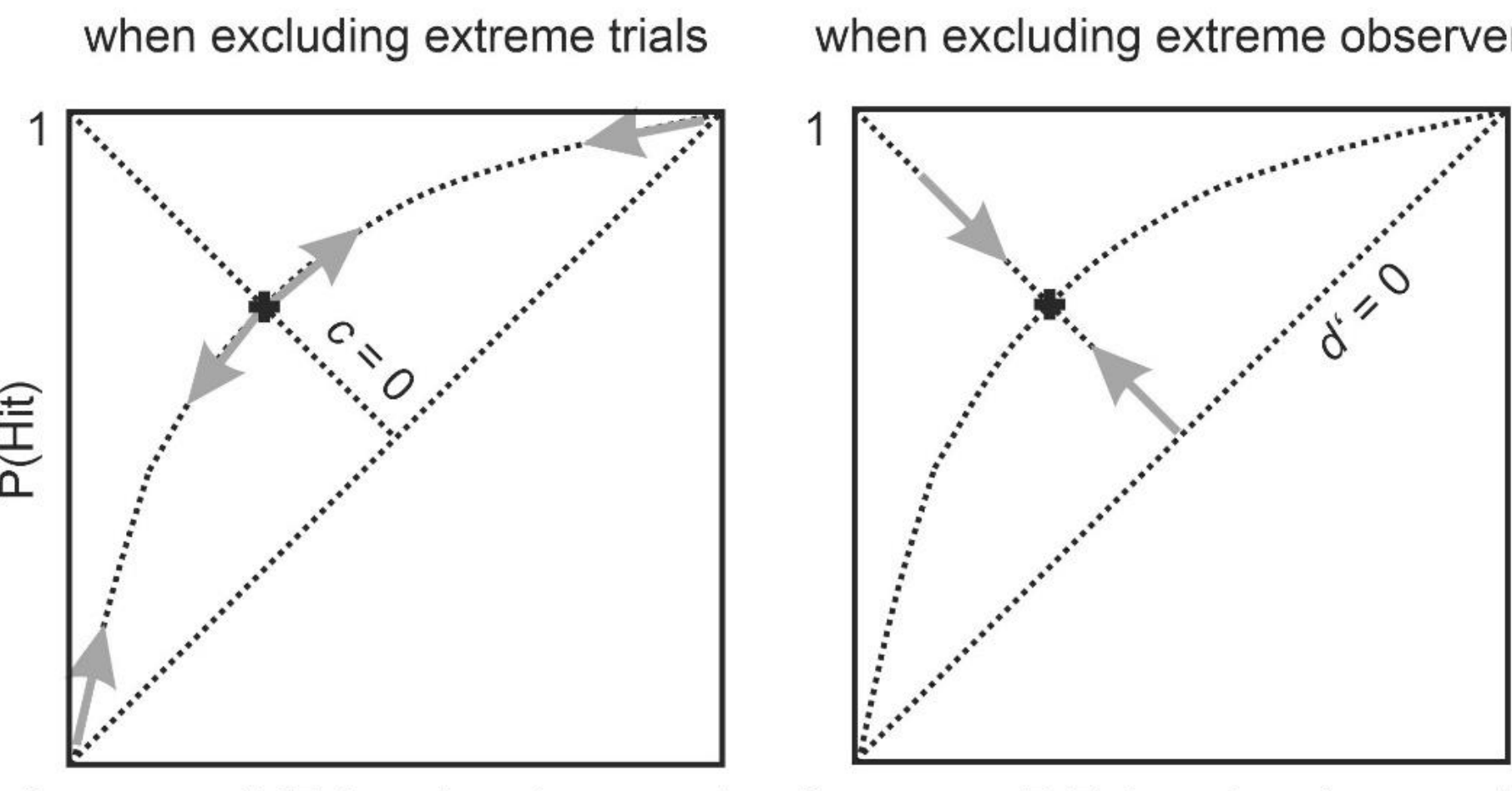

*Figure 3: Direction of squeeze effects in discrimination tasks when regression to the mean occurs on the basis of selective analysis. Left panel: When selectively analyzing trials with very low or very high visibility ratings, squeeze effects are directed away from extreme ratings and towards the center of the trial distribution for each of the stimulus alternatives. Right panel: When selectively analyzing observers with very low or very high performance, squeeze effects are directed away from extreme performances and towards the true underlying sensitivity of the observer distribution.*

What is true for discrimination tasks (2YN and 2AFC) also holds for detection and “liminal prime” tasks. In a detection experiment, we compare a noise distribution *N* (on the left of the decision axis) and a signal-plus-noise distribution *S* (to the right), with their normalized distance $d'$ defining the observer’s detection sensitivity. Now hits and false alarms are defined as correct responses to *S* and incorrect responses to *N*, respectively. Again, visibility ratings cut this detection space into vertical stripes, and the overall sensitivity determines every point on the detection ROC, the low-visibility and the high-visibility ratings alike. When selecting detection trials with extreme ratings, squeeze effects are directed from the ends of the curve towards the center (in detection, the central rating categories would normally not be regarded as interesting candidates for selection). In contrast, when selecting observers with extreme performance, squeeze effects are directed towards the true underlying sensitivity just as in the discrimination task. -- In the liminal-prime paradigm (Avneon & Lamy, 2018; Bachmann & Francis, 2014; van den Bussche et al., 2013), the same weak stimulus is presented over and over again, and the variance in visibility ratings is thought to arise from spontaneous fluctuations in visibility. Because of the lack of a comparison condition (e.g., an S2 or N), this task design is principally unable to separate sensitivity and response criteria. In such a study, which is no longer experimental but merely correlational, the outcome may be determined exclusively by measurement noise. [5]

---

[5] Another problem of the liminal-prime paradigm is that many sources of spontaneous fluctuations can be expected to affect indirect measures as much as direct measures. For instance, fluctuations in bottom-up signal strength should affect all information processing downstream, and top-

So far we have argued from signal detection theory's standard model for constructing ROCs where discrimination responses and visibility ratings share the same decision axis, but it should be acknowledged that the details depend on the multivariate geometry of the decision space (Hellmann et al., 2023; Locke et al., 2022; Malejka & Bröder, 2019; Rausch et al., 2021; Zehetleitner & Rausch, 2013) and on the way criteria are placed for different types of direct measures (Macmillan, 1986; Maniscalco & Lau, 2012, 2014; Maniscalco et al., 2016). Any two measures that are appreciably different in their criterion contents can be expected to have different decision axes, even though these can be close to each other (e.g., highly correlated) in multivariate space. For instance, it has been shown that visibility judgments and decision confidence do not lead to identical results (Rausch & Zehetleitner, 2016; Rausch et al., 2015). Now a common validity check for PAS ratings is to show that as sensitivity increases, ratings systematically shift from higher to lower categories, and of course that is what they do. This behavior implies that the PAS ratings do indeed have some life of their own and that they use a decision axis not entirely identical to that of the forced-choice measure. This is not necessarily a good thing, though: possibly it merely reflects the PAS's tendency to ask for overall detectability and confidence regarding "the stimulus" as a whole and not for the actual discrimination of the critical feature (Schmidt & Biafora, 2024). Studies that compared discrimination and PAS ratings have sometimes found a great deal of convergence between the measures (Kiefer et al., 2023), but at other times striking discrepancies, especially when looking at the behavior of single observers (Biafora & Schmidt, 2022).

## 5. The stripe-by-stripe fallacy

We have seen that in discrimination as well as detection tasks, measurements at different response criteria (like the lowest PAS rating) can only be interpreted if the overall sensitivity is known. But many studies have attempted to measure sensitivity separately for different rating categories without realizing that sensitivity is a property of the *distribution* of trials, not of single trials. Many computational artifacts can arise from calculating indices of sensitivity separately for different rating categories. Again, consider Fig. 2a where the rating categories are numbered from negative to positive, and let $A_i$ equal the percentage of correct responses in rating category $i$. When $d' = 0$, the S1 and S2 distributions are identical, and $A_i = 0.5$ in each rating category (chance level guessing). But whenever $d'$ is not zero, the lowest rating categories will always have lower accuracy than the highest rating categories, just like we have seen above in the data from Biafora and Schmidt (2022). This occurs by necessity because the lowest rating categories will come from the area of largest overlap between the S1 and S2 distributions, while the highest rating categories come from the tails where misses and false alarms are rare. However, it would be a mistake to interpret any of those seemingly different accuracy levels in isolation: none of them is free to vary because down modulations of visual attention have been shown to affect response priming (Schmidt & Seydell, 2008; Schmidt & Schmidt, 2010).

they are all jointly determined by the degree of overlap between S1 and S2 (i.e., by sensitivity). By the same argument, Stein et al. (2016; see also Fahrenfort et al., 2024) have shown that is a mistake to calculate *d′* (or any other sensitivity statistic) separately for each rating category, even though this is frequently seen in the literature (e.g., Peremen & Lamy, 2014; Soto et al., 2011; Wixted et al., 2015; see Soto & Silvanto, 2016, for a reply to some of the criticism). Those statistics will be systematically too small for the rating categories in the center of overlap and systematically too large for the rating categories in the periphery.

Generally, a simple detection or discrimination experiment can only have one underlying sensitivity because there is only one normalized distance between two distributions of internal evidence, and therefore all the seemingly different values must be a function of response criterion. This principle applies to all kinds of performance measures, like percentage of correct responses, various sensitivity statistics like *d′* or *H-F*, and psychometric functions. For example, King and Dehaene (2014), after providing a detailed and correct analysis of the stripe-by-stripe problem, nevertheless report an experiment where participants had to decide between two numerals, with various degrees of intermediate morphs along the decision axis. This experiment would normally yield a single psychometric function for each observer. Yet they report *several* such functions, one for each visibility rating, and conclude that high-visibility stimuli are characterized by steeper psychometric functions than are low-visibility stimuli. In fact, the different slopes are a result of the stripe-by-stripe fallacy: by conditioning on the visibility rating, we select subsets at different response criteria, but all from the same underlying sensitivity function. As before, there is only one underlying psychometric function, and the apparent split into several is a statistical artifact.

## 6. General Discussion: From post-hoc selection to functional dissociations

The problem of unconscious cognition is a true classic of experimental psychology. Even though the field possesses an unbroken research tradition dating back into the late 19th century, it was marred by boom-and-bust cycles in which a resurgence of interest in the topic first led to an attrition of methodological criteria and then was ended by seminal review papers that called the validity of the entire field into question (e.g., Eriksen, 1960; Holender, 1986). Both times this happened the field collapsed and fell into disregard, if not disrepute, for a decade or more. If there is any lesson from the history of our field, it is this: Convincing each other that our methods are “recommended” or “good practice” is just not good enough. The ambition must be to convince hardcore skeptics, past, present and future, that our methods are valid and can stand the test of time. That is only possibly by using strict standards of measurement, experimental design, and psychophysical modeling, and by avoiding well-known statistical artifacts that would bias our conclusions towards the “wild hypothesis”. If methodological standards in unconscious cognition systematically fall short of what is expected in adjacent fields like masking or response time research, and if we neglect basic principles of statistics, psychophysics, and psychometrics, then our field is increasingly in danger of encapsulation and ceases to be interesting, relevant, or even respectable to others.

Stockart et al.'s attempt to reach a consensus among researchers from various schools and methodological preferences leads to many recommendations that we share. But in its current form, the project condones the widespread practice of post-hoc selection on the basis of multiple responses per trial. But post-hoc selection is not a valid technique for isolating trials without awareness. At the heart of the problem is a sampling fallacy: selection on the basis of a *sample* does not change the properties of the *population* on which the sample is based (Schmidt, 2015). Even though subjective visibility may fluctuate from trial to trial, and even if a rating scale validly captures those fluctuations, an observer's sensitivity can still be calculated *across* trials and may easily be too high to plausibly argue for unconscious cognition. If sensitivity is represented by the shape of the ROC curve, then the lowest visibility ratings simply represent the points on the curve that are associated with the largest subjective uncertainty, but still those points are part of the very same ROC as all the higher visibility ratings. In discrimination tasks, they lie right in the center of the curve and could be obtained by interpolation even if low PAS ratings never factually occurred – after all, *any* ROC has to pass the minor diagonal.

Relatedly, selecting any particular visibility category and calculating separate sensitivity statistics for that subsample is a miscalculation of psychophysical measures that results in values that are too low when subjective visibility is low and too high when subjective visibility is high (the stripe-by-stripe fallacy; Stein et al., 2016). Unreliability in the rating scale contributes to these problems by introducing regression to the mean, which leads to an underestimation of visibility at low ratings and an overestimation at high ratings (Shanks, 2017; Shanks et al., 2021). Regression to the mean particularly threatens the validity of low-visibility ratings, which capitalize on the noise in the measurement (Schmidt, 2015). While this problem mainly affects which parts of the ROC enter the post-hoc selection, it is the selection process itself that is invalid. Whenever visibility ratings come from a stimulus condition with an ROC that is too steep, then sensitivity in that condition was too high even if some of its trials received a low rating. Note that the fundamental problem here is not one of objective versus subjective measures, but one of trial-by-trial versus condition-wide measures: If we used a subjective measure instead of objective $d'$ to evaluate an entire stimulus condition, that measure may speak against unconscious processing as well.

All these problems apply not only to discrimination tasks, but also to detection tasks (where rating categories are just arranged differently along the detection ROC) and the "liminal-prime" paradigm (which cannot tell signal from noise in the first place and may be driven entirely by measurement noise). They arise from a fundamental flaw in the research approach: instead of controlling visibility experimentally, one first leaves it to chance and then turns one of the dependent variables (visibility rating) into an independent factor for the analysis of another dependent variable (e.g., response time effect). Using a dependent variable as if it was experimentally controlled is always a flirt with disaster in data analysis: it often leads to unexpected artifacts as the orderly and balanced experimental design is degraded into an unbalanced correlational design, with its usual share of unforeseen conditional probabilities and potential confounds. But as soon as visibility is under experimental control, many of the potential biases and restrictions of visibility measures cease to be so threatening. For instance, Skóra et al. (2021) argue that *absolute* PAS ratings are affected by response bias but that *relative* ratings are not: PAS profiles may change in absolute magnitude but can still be compared across conditions. Such data indicate that there is a useful function for PAS-like rating scales if the rating profile is compared across experimental conditions. In contrast, if we wanted to use those ratings as a basis for post-hoc selection, any such shift in absolute magnitude would distort the results.

Functional dissociations between direct and indirect measures solve most if not all of those problems. First, there are many dissociation patterns that neither require nor benefit from null sensitivity in the direct measure (they are literally "beyond the zero-awareness criterion", Schmidt, 2007). Second, they all utilize parametric variations of stimulus parameters like stimulus intensity or prime-mask SOA, providing research designs with internal validity ensuring that functions are orderly, well-behaved, and measured with precision. Third, they all trace out meaningful functional relationships that are as relevant within our field as they are in more specialized areas (e.g., priming and masking functions).

The most basic type of functional dissociation is an invariance dissociation (or its special case, the simple dissociation) where one of the measures is shown to remain invariant (or close to zero) as the other one is experimentally varied. For instance, response priming increases with prime-target SOA even as prime discrimination performance is invariant at chance level (Vorberg et al., 2003, Exp. 1; F. Schmidt & Schmidt, 2010). Conversely, simple reaction times to the prime remain invariant as the strength of metacontrast masking is experimentally varied (Fehrer & Raab, 1962), and the time-course of response priming remains invariant no matter whether the ability to discriminate the prime is low, high, increasing, or decreasing (Vorberg et al., 2003, Exp. 2). Because both priming and masking functions are under full experimental control and no post-hoc selection is taking place, neither sampling bias nor regression to the mean play any role. Invariance and simple dissociations are, however, dependent upon the assumption that the invariant measure is exhaustively valid and reliable with respect to the critical feature: otherwise, the invariance could simply arise from limited sensitivity to the relevant information in the prime. Sensitivity dissociations as demanded by Meyen et al. (2022) are more flexible: they do not require invariance in any of the measures but only that the indirect measure systematically exceeds the direct one (after suitable standardization). Still, however, sensitivity dissociations require an assumption that the direct measure is in principle at least as sensitive to the critical feature as the indirect measure (Schmidt & Vorberg, 2006). Finally, the most informative data patterns come from double dissociations: The demonstration that the same experimental manipulation leads to opposite effects on direct and indirect variables directly shows that they cannot both depend monotonically on the same single source of conscious information. Double dissociation patterns also require the mildest measurement assumptions, because they do not require any assumptions of relative sensitivity, exclusiveness, or exhaustive reliability (see mathematical proofs in Schmidt & Vorberg, 2006; Biafora & Schmidt, 2020, 2022).

A major incentive for using post-hoc selection on the basis of multitasks is certainly the desire for a seemingly foolproof method to obtain trials with unconscious primes, even if only in a fraction of the experiment. In contrast, past researchers working on simple point dissociations were forced to search for a tiny goldilocks zone of experimental conditions that seemed to allow for just the right amount of priming under just the right amount of masking, often exploiting low measurement precision to argue that discrimination performance for the critical stimulus was not statistically different from chance (Stein et al., 2024). In such a paradigm there is always an incentive to accept an underpowered measurement to use the absence of significant direct effects as an argument for unawareness. One advantage of functional dissociations is that they always reward measurement precision because they are usually not confined to a specific value of the direct measure.

In their study of subjective percepts in metacontrast-masked primes, Koster et al. (2020) have demonstrated a number of functional double dissociations among several different subjective percepts, all of them potential measures of awareness. Findings such as that make it clear that there cannot be a single measure to capture all of "consciousness": There are many different aspects of visual awareness that should be studied in their own right (Fazekas & Overgaard, 2018; Klein & Hohwy, 2015; Seth et al., 2008), and conscious percepts like perceived brightness, color, or motion require their own specialized theoretical frameworks anyway (Irvine, 2017; Schmidt & Biafora, 2024). Functional dissociations offer a research program for seriously studying various aspects of visual awareness under different task demands. They allow us to unfold the traditional dissociation paradigm by utilizing the entire *D-I* space, not just single points on a single line. Making use of the full range of functional dissociations allows us to conduct successful experiments in a reliable and predictable way – not just by sheer luck or by capitalizing on chance, but by the planful study of theoretically meaningful relationships between experimentally controlled variables.